\documentclass[aps,prl,twocolumn,floatfix,superscriptaddress,showpacs,amsfonts,amssymb,amsmath]{revtex4}

\usepackage{graphicx}% Include figure files
\usepackage{dcolumn}% Align table columns on decimal point
\usepackage{bm}% bold math
\usepackage[dvips]{color}

\begin{document}
\renewcommand{\r}[1]{\big<\, r^{#1}\, \big>_{\rm tr}}

\title{Pair Decay Width of the Hoyle State and its Role for Stellar Carbon Production}

\author{M.~\surname{Chernykh}}
\affiliation{Institut f\"ur Kernphysik, Technische Universit\"at
Darmstadt, D-64289 Darmstadt, Germany}
\author{H.~\surname{Feldmeier}}
\affiliation{Gesellschaft f\"ur Schwerionenforschung (GSI), D-64291
Darmstadt, Germany} \affiliation{Yukawa Institute for Theoretical
Physics, Kyoto, Japan}
\author{T.~\surname{Neff}}
\affiliation{Gesellschaft f\"ur Schwerionenforschung (GSI), D-64291
Darmstadt, Germany}
\author{P.~\surname{von Neumann-Cosel}}
\email{vnc@ikp.tu-darmstadt.de} \affiliation{Institut f\"ur
Kernphysik, Technische Universit\"at Darmstadt, 64289 Darmstadt,
Germany}
\author{A.~\surname{Richter}}
\affiliation{Institut f\"ur Kernphysik, Technische Universit\"at
Darmstadt, D-64289 Darmstadt, Germany} \affiliation{ECT*, Villa
Tambosi, I-38100 Villazzano (Trento), Italy}

\date{\today}

\begin{abstract}
The pair decay width of the first excited $0^+$ state in $^{12}$C (the
Hoyle state) is deduced from a novel analysis of the world data on
inelastic electron scattering covering a wide momentum transfer range,
thereby resolving previous discrepancies. The extracted value
$\Gamma_{\pi} = (62.3 \pm 2.0)$~$\mu$eV is independently confirmed by
new data at low momentum transfers measured at the \mbox{S-DALINAC} and
reduces the uncertainty of the literature values by more than a factor
of three.
%
%Electron scattering off the first excited $0^+$ state in $^{12}$C (the
%Hoyle state) has been performed at low momentum transfers at the
%\mbox{S-DALINAC}. The new data together with a novel model-independent
%analysis of the world data set covering a wide momentum transfer range
%result in a highly improved transition charge density from which a pair
%decay width $\Gamma_{\pi} = (62.3 \pm 2.0)$~$\mu$eV of the Hoyle state
%was extracted reducing the uncertainty of the literature values by more
%than a factor of three.
%
A precise knowledge of $\Gamma_\pi$ is mandatory for quantitative
studies of some key issues in the modeling of supernovae and of
asymptotic giant branch stars, the most likely site of the slow-neutron
nucleosynthesis process.

\end{abstract}


\maketitle

%INTRODUCTION:

The production of the most abundant stable carbon isotope $^{12}$C in
the center of stars proceeds through the fusion of three $\alpha$
particles (the triple-$\alpha$ process). Its reaction rate is of
critical significance for a variety of key problems of nuclear
astrophysics \cite{ref:Fynbo2005} like the elemental abundance in the
universe \cite{ref:Wallerstein1997}, the size of the iron core in
massive stars determining the properties of supernova explosions
\cite{ref:Austin2005,ref:Tur2007}, the dynamics of asymptotic giant
branch (AGB) stars \cite{ref:Herwig2006}, the site of the main
slow-neutron capture process ($s$-process) of heavy elements
\cite{ref:Kappeler1998}, or the weak $s$-process in massive stars
\cite{ref:Tur2009}.

Under most astrophysical conditions the reaction exclusively proceeds
through a scattering resonance of three $\alpha$-particles that
represents an excited state in $^{12}$C with quantum numbers $J^\pi =
0^+$ at an excitation energy $E_x = 7.654$ MeV (the so-called Hoyle
state \cite{ref:Hoyle1954}), which then decays to the stable ground
state. The reaction rate of the resonant triple-$\alpha$ process is
proportional to the radiative decay width $\Gamma_{rad}$ of the Hoyle
state. Although this state is experimentally studied for more than 50
years, at present $\Gamma_{rad}$ is known with an uncertainty of about
$\pm$12\% only. However, an accuracy of about $\pm 5$\% is requested
for applications in astrophysics \cite{ref:Austin2005,ref:Herwig2006}.

Experimentally the radiative width cannot be accessed directly but
is determined as a product of quantities
%---------------------------------------------------------
\begin{equation}
\Gamma_{rad} = \Gamma_{\gamma} + \Gamma_{\pi} =
\frac{\Gamma_{\gamma} + \Gamma_{\pi}}{\Gamma}\cdot\,
\frac{\Gamma}{\Gamma_{\pi}}\cdot\,\Gamma_{\pi}
 \label{eq:RadiationWidth}
\end{equation}
%---------------------------------------------------------
measured in different experiments. Here, $\Gamma = \Gamma_{\alpha} +
\Gamma_{\gamma} + \Gamma_{\pi}$ is the total decay width taking into
account $\alpha$, $\gamma$ and $e^{\pm}$ decays of the Hoyle state.
Whereas the first quantity on the right-hand side of
Eq.~(\ref{eq:RadiationWidth}) has been precisely determined with an
uncertainty of $\pm$2.7\% (see Ref.~\cite{ref:Markham1976} and refs.\
therein), the branching ratio $\Gamma_{\pi}/\Gamma$ is known with an
error of $\pm$9.2\% only (see Ref.~\cite{ref:Alburger1977} and refs.\
therein). However, a new measurement of $\Gamma_{\pi}/\Gamma$ with an
expected precision of about $\pm$5\% is currently
underway~\cite{ref:Austinprivcomm}.

The pair decay width $\Gamma_{\pi}$ in Eq.~(\ref{eq:RadiationWidth})
can be determined by inelastic electron scattering
%.
%
through the relation~\cite{ref:Wilkinson1969}
\begin{equation}
\Gamma_\pi = C(Z,E_x) \frac{8 \alpha^2}{27 \pi} \frac{B(E_x)}{(\hbar c)^4}
F(E_x) \r{2},
\label{eq:pairwidth}
\end{equation}
where $F(E_x) = (0.5E_x - m_ec^2)^3 (0.5E_x + m_ec^2)^2$, $\alpha$
denotes the fine structure constant and $\r{2}$ the matrix element of
the monopole transition. The energy-dependent correction term $B(E_x)$
is given in Tab.~1 of Ref.~\cite{ref:Wilkinson1969} (0.898 for the
present case). The influence of the nuclear Coulomb field is taken into
account by the factor $C(Z,E_x)$ and amounts to 1.013 for the Hoyle
state. The reduced $E0$ transition probability is the largest known for
excitation of a single state \cite{ref:kibedi2005} and exhausts about
7.5\% of the energy-weighted sum rule.

Two values were extracted from measurements at low momentum transfers
$q$ with a so-called 'model-independent' analysis
\cite{ref:Crannell1967,ref:Strehl1970} in the distorted wave Born
approximation (DWBA) explained in detail below. They agree well with
each other and an averaged value $60.5 \pm 3.9$~$\mu$eV is quoted in
Ref.~\cite{ref:Ajzenberg-Selove1990}. Alternatively, a Fourier-Bessel
analysis~\cite{ref:Heisenberg1981} of the transition form factor
including data over a wide momentum transfer range has recently been
done by Crannell {\it et al.}~\cite{ref:Crannell2005}. The extracted
value of $\Gamma_\pi = 52.0 \pm 1.4$~$\mu$eV has an error of 2.7\%
only, but the low-$q$ result deviates by about 6$\sigma$.

To resolve this discrepancy we have performed a new high-resolution
electron scattering experiment at low momentum transfer and
reinvestigated both approaches. A novel model-independent ansatz was
developed to analyze the global form factor. This provides not only a
precise determination of the pair width $\Gamma_{\pi}$ but also an
important test of current models for the structure of the Hoyle state
\cite{ref:Chernykh2007}, which represents a prime example of an
$\alpha$-cluster state predicted to possess the features of a
low-density gas of $\alpha$ particles resembling a Bose-Einstein
condensate \cite{ref:Tohsaki2001}. As shown in
Ref.~\cite{ref:Chernykh2007}, state-of-the-art models describe the
$(e,e')$ form factor quite well over a broad momentum transfer range
but predict a monopole matrix element (directly related to
$\Gamma_\pi$) about 20\% too large (cf.\ Table~I in
\cite{ref:Chernykh2007}). The present very precise results confirm this
discrepancy and provide new insight into its origin.

\begin{figure}[tbh!]
\includegraphics[width=7.5cm]{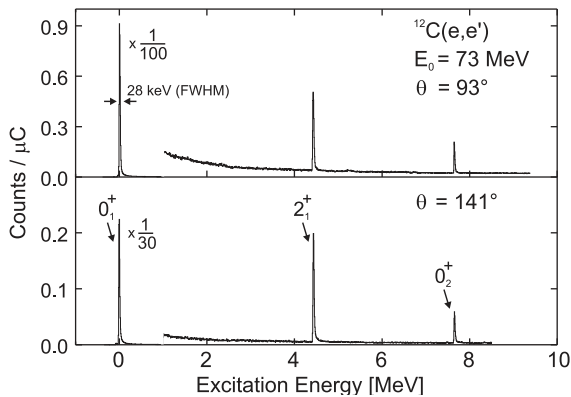}
\caption{\label{fig:C12eeRawSpeAll} Spectra of the
$^{12}$C(e,e$^{\prime}$) reaction measured at a beam energy
$E_{0} = 73$~MeV and scattering angles $\theta$ = 93$^{\circ}$
(top) and $\theta$ = 141$^{\circ}$ (bottom).}
\end{figure}

The experiment was carried out at the high energy-resolution magnetic
spectrometer~\cite{ref:lenhardt2006} of the Darmstadt superconducting
electron linear accelerator S-DALINAC. Excitation energy spectra were
taken at initial electron energies between 29~MeV and 78~MeV and
scattering angles from 69$^{\circ}$ to 141$^{\circ}$ with beam currents
of about 1~$\mu$A. The momentum transfer range for the excitation of
the Hoyle state varied thus between 0.21~fm$^{-1}$ and 0.66~fm$^{-1}$.
A self-supporting carbon target (98.9\% $^{12}$C) with an areal density
of 6.4~mg/cm$^2$ was used. In dispersion-matching mode an energy
resolution $\Delta$E~$\approx$~28~keV (full width at half maximum,
FWHM) was achieved. Examples of spectra at a beam energy of 73~MeV are
presented in Fig.~\ref{fig:C12eeRawSpeAll}. The peaks correspond to the
elastic line, the $2^{+}_{1}$ state and the Hoyle state in $^{12}$C.
%
%fffffffffffffffffffffffffffffffffffffffffffffffffffffffffffffff
%fffffffffffffffffffffffffffffffffffffffffffffffffffffffffffffff

As the form factor is the Fourier transform of the density, it
corresponds to a cross section calculated in plane wave Born
approximation (PWBA). The measured cross section, however, corresponds
to the distorted wave Born approximation (DWBA). Therefore, we deduce
the square of the experimental form factor - the differential cross
section divided by the Mott cross section - at the measured momentum
transfer $q_i$ and electron beam energy $E_{0i}$ by
%-------------------------------------------------------------------------
\begin{equation}\label{eq:FF-exp}
\left|F^{\rm exp}_{\rm tr}(q_i)\right|^2= \frac{4\pi}{Z^2} B(C0,q_i,E_{0i})\
\frac{B^{\rm PWBA}(q_i)}{B^{\rm DWBA}(q_i,E_{0i})}\ ,
\end{equation}
%-------------------------------------------------------------------------
where $Z$ is the number of protons and $B(C0,q_i,E_{0i})$ denotes the
reduced transition probability, while $B^{\rm PWBA}(q_i)$ and $B^{\rm
DWBA}(q_i,E_{0i})$ are reduced transition probabilities calculated
within plane wave and distorted wave Born approximation, respectively.
This so-called PWBA transformation allows to relate cross sections
measured at different $q_i$ and $E_{0i}$ directly to the form factor or
corresponding transition density.
%It is a good approximation if the momentum
%transfer measured at large distance does not differ too much from the
%momentum transfer close to the nucleus.
The data of the different experiments should then collapse into a
single line for $\left|F^{\rm exp}_{\rm tr}(q_i)\right|^2$. For the
present measurements typical transformation factors are around $0.85$.

The factors $B^{\rm PWBA}(q_i)/B^{\rm DWBA}(q_i,E_{0i})$ in
Eq.~(\ref{eq:FF-exp}) are computed with the code of Heisenberg and
Blok~\cite{ref:Heisenberg1983} in an iterative procedure starting with
a the transition density $\rho_{\rm tr}(r)$ taken from the models
discussed in Ref.~\cite{ref:Chernykh2007}, e.g.\ the Fermionic
Molecular Dynamics (FMD) model \cite{ref:Roth2004}. The resulting form
factor is transformed into an improved transition density that enters
in a second step into the PWBA transformation in Eq.~(\ref{eq:FF-exp})
etc. Convergence is reached after three steps.

The transition form factor $F_{\rm tr}(q)$ of a monopole transition is
related to the transition density $\rho_{\rm tr}(r)$ by
%----------------------------------------------------------------------
\begin{eqnarray}\label{eq:FF-rho}
F_{\rm tr}(q) & = & \frac{4\pi}{Z}\int_0^\infty \rho_{\rm tr}(r)\ j_0(qr)\ r^2\ dr
\\ \nonumber
& = & \frac{1}{Z} \sum_{\lambda=1}^\infty \frac{(-1)^\lambda}{(2\lambda+1)!} \
  q^{2\lambda}\  \r{2\lambda} \ ,
\end{eqnarray}
%-----------------------------------------------------------------------
where $\rho_{\rm tr}(r)=\big<0^+_1\big|\hat{\rho}(\vec{r})\big|0^+_2\big>$
is the matrix element of the charge density operator $\hat{\rho}(\vec{r})$
between the ground state and the Hoyle state and the operator
$\hat{Z}=\int \hat{\rho}(\vec{r})\,d^3r$ counts the number of protons.

Expansion of the Bessel function $j_0(qr)$ for low $q$, i.e. $q^2 \r{2}
\ll 1$, shows that the form factor in Eq.~(\ref{eq:FF-rho}) is governed
by the lowest moments $\r{2}$ and $\r{4}$. Hence precise data at low
$q$ are important for an accurate determination of the pair width
$\Gamma_\pi$ which is proportional to $\r{2}^2$, cf.\
Eq.~(\ref{eq:pairwidth}). But as we shall show later the power
expansion around $q=0$ is very sensitive to statistical and systematic
errors. Therefore we combine our new results with the previous world
data set \cite{ref:Crannell2005} and perform a model-independent
analysis of the $0^+_1 \rightarrow 0^+_2$ transition form factor by a
global fit to all data.

In the present analysis we use the model-independent ansatz
%------------------------------------------------------------------------
\begin{equation}\label{eq:FFfit}
F_{\rm tr}(q)=\frac{1}{Z}\ {\rm e}^{-\frac{1}{2}(b\,q)^2}\
                               \sum_{n=1}^{n_{\rm max}} c_n\ (b\,q)^{2n}
\end{equation}
%------------------------------------------------------------------------
which respects the condition $F_{\rm tr}(q=0)=0$ and the fact
that $j_0(qr)$ contains only even powers of $q$.
%and thus $F_{\rm tr}(q)$ is actually a function of $q^2$.
The fit parameters are $b$ and $c_n$, where $b$ plays the role of a
length scale. The matrix element $\r{2}=|6 c_1 b^2|$ is simply given by
$c_1$ and $b$.

The transition density $\rho_{\rm tr}(r)$ corresponding to the ansatz (\ref{eq:FFfit})
is given by
%------------------------------------------------------------------------
\begin{equation}\label{eq:rhofit}
\rho_{\rm tr}(r)=\frac{1}{b^3}\ {\rm e}^{ -\frac{1}{2}\left(\frac{r}{b}\right)^2}\
   \sum_{n=0}^{n_{\rm max}} d_n \left(\frac{r}{b}\right)^{2n} \ ,
\end{equation}
%------------------------------------------------------------------------
where the relation between the coefficients $c_n$ and $d_n$ can be
calculated by the inverse transformation of Eq.~(\ref{eq:FF-rho}).
%, and
%$\rho_{\rm tr}(r)$ is determined only up to a phase which we choose
%such that $\r{2}$ is positive.
For the Hoyle state measurements of
$|F^{\rm exp}_{\rm tr}(q_i)|^2$ exist up to $q_{\rm max} =
3.1$~fm$^{-1}$ so that structures in $\rho_{\rm tr}(r)$ can be resolved
on a scale of $\pi/q_{\rm max}\simeq1$~fm.

\begin{figure}[tbh]
\includegraphics[width=7.5cm]{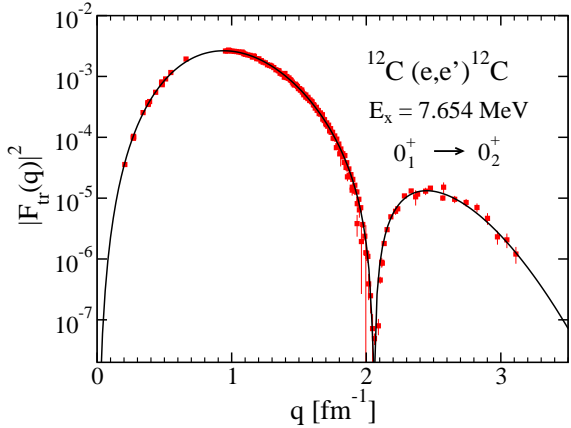}
\caption{\label{fig:FFglobal} (color online). Squared form factor for the transition from
the ground state in $^{12}$C to the Hoyle state extracted by the
model-independent analysis (solid line) described in the text in comparison to the
experimental data \cite{ref:Strehl1970,ref:Crannell2005} transformed
according to Eq.~(\ref{eq:FF-exp}).}
%Data with $|F^{\rm exp}_{\rm tr}|^2 \leq 10^{-6}$ are excluded from the fit.}
\end{figure}

%For an accurate extraction of the transition density the data should
%cover a broad range in momentum transfer. This is the case
%\cite{ref:Strehl1970,ref:Crannell2005,ref:CrannellPrivComm} that have also been utilized to
%elucidate its structure in Ref.~\cite{ref:Chernykh2007}.
%
Due to the Coulomb distortion of the in- and outgoing scattering states
the cross section remains finite at momentum transfers where the form
factor goes through zero and hence the uncertainty in deducing $|F_{\rm
tr}^{\rm exp}(q)|$ according to the PWBA transformation
Eq.~(\ref{eq:FF-exp}) is large around $q=2.1~\rm{fm}^{-1}$, see
Fig.~\ref{fig:FFglobal}. Therefore only data $|F_{\rm tr}^{\rm
exp}(q)|^2 > 10^{-6}$ are considered. Variation of the maximum power of
the polynomial in Eq.~(\ref{eq:FFfit}) shows that $n_{max} = 5$ is
sufficient for the present analysis.
%
%fffffffffffffffffffffffffffffffffffffffffffffffffffffffffffffffffffff
%fffffffffffffffffffffffffffffffffffffffffffffffffffffffffffffffffffff

Figure \ref{fig:FFglobal} shows the resulting global fit. It describes
the data well over the whole $q$ range of measured momentum transfers
including the minimum region, where data were excluded from the fit.
The following transition radii are extracted: $\r{2}= (5.47 \pm
0.09)$~fm$^2$ and $\r{4}= (115 \pm 8)$~fm$^4$. The uncertainties are
estimated by varying the data base because they are dominated by
systematic errors in the data sets. The accuracy is essentially limited
by the experimental uncertainties at $q > 1.7$~fm$^{-1}$ where the
cross sections are small. This might look surprising because the
transition radius is given by the limit $q\rightarrow 0$. However, we
have reached an accuracy where seemingly small contributions become
relevant and furthermore the form factor has to be a smooth function so
that information from higher $q$ values may influence the low-$q$
behavior to a certain extent.

%ffffffffffffffffffffffffffffffffffffffffffffffffffffffffffffffffffffffffffff
\begin{figure}[tbh]
\includegraphics[width=7cm]{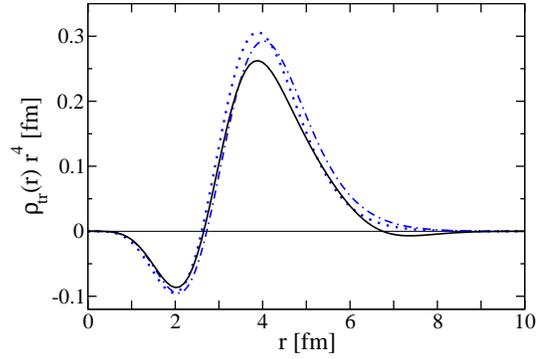}
\caption{\label{fig:rhofit} (color online). Transition charge density
corresponding to the form factor displayed in Fig.~\ref{fig:FFglobal}
%of the transition from the ground state in $^{12}$C to the Hoyle state
multiplied by $r^4$
%obtained in a global model independent fit to all
%available data with $|F^{\rm exp}_{\rm tr}|^2 \geq 10^{-6}$
(full line) in
comparison to the theoretical predictions \cite{ref:Chernykh2007}
of FMD (dotted line) and an $\alpha$ particle gas (dashed-dotted line).
}
\end{figure}
%ffffffffffffffffffffffffffffffffffffffffffffffffffffffffffffffffffffffffffffffffff

In Fig.~\ref{fig:rhofit} we display $\rho_{\rm tr}(r)r^4$ to show that
$\r{2}$, which is $4\pi$ times the integral over this function, is very sensitive to
contributions beyond $r=4$~fm where the ground state density is very small but
the density of the Hoyle state is not, see Ref.~\cite{ref:Chernykh2007}.
The transitions densities obtained in Fermionic Molecular Dynamics and
BEC (gas of $\alpha$-particles) are somewhat too large beyond 4~fm
indicating that the calculated charge density of the Hoyle state is overestimated
at large distance.
Although the deviation does not seem to be big the transition radius $\r{2}^{calc}$ is
for both models about 20\% too large because negative and positive contributions
cancel up to $r \approx 3.5$~fm and only the region beyond 3.5~fm matters.

% Vergleich mit Crannell 2005
%
Crannell {\it et al.}~\cite{ref:Crannell2005} used the same data,
except for the seven new data points, and performed a Fourier Bessel
(FB) analysis \cite{ref:Heisenberg1981}. Figure~\ref{fig:rhofit}
illustrates that beyond their adopted cut-off radius $R_c \approx
8.5$~fm the transition density vanishes. They concluded however a
significantly smaller $\Gamma_\pi$. To understand this discrepancy we
take our fit as an approximation of the experimental form factor and
read off the FB expansion coefficients at $q_\nu = \nu \pi / R_c$ for
$\nu =1$ up to $\nu_{max}=8$. The resulting transition density has
$\r{2}^{\rm FB}=4.99~\rm{fm}^2$ or $\Gamma_\pi^{\rm FB}=51~\mu$eV in
accordance with the result of \cite{ref:Crannell2005}. In order to
reproduce our own result within a FB analysis we have to go up to at
least $\nu_{max}=10$  which corresponds to $q_{max}=3.7~\rm{fm}^{-1}$
where no data exist.
The same holds true if we repeat the analysis without the S-DALINAC
data.
In general one should keep in mind that due to the oscillatory behavior
of the Bessel functions a FB expansion can only reproduce the tail
region well, if $\nu_{max}$ and $q_{max}$ are sufficiently large.

% Low q expansion

The Taylor expansion of the form factor around $q=0$ offers itself a
model-independent way to deduce the transition radius $\r{2}$ which is
the leading coefficient in the expansion. To study this often employed
approach we deduce from Eq.~(\ref{eq:FF-rho})
\newcommand{\lowq}{{{\rm low}\,q}}
%------------------------------------------------------------------------
\begin{equation}\label{eq:fit}
-6 Z \frac{F_{\rm tr}(q)}{q^2} =
\r{2}^\lowq -\frac{q^2}{20}\, \r{4}^\lowq + \dots
%+\frac{q^4}{840}  \,  x_3 \left(\r{2}^\lowq\right)^3
%-\frac{q^6}{60480}\,  x_4 \left(\r{2}^\lowq\right)^4
\end{equation}
%------------------------------------------------------------------------
which is used to fit the low-$q$ data measured at the S-DALINAC.
%at seven values of  $q^2$ between 0.043 and 0.437 fm$^{-2}$.
The quantities $\r{2}^\lowq$ and $\r{4}^\lowq$ are taken as free
parameters. Higher powers up to $q^8$ are included in the fit such that the
last term in Eq.~(\ref{eq:fit}) contributes less than 1\% of the first
term. Because of the limited number of data points we approximated the
higher transition matrix elements by $(\r{2}^\lowq)^n \cdot x_n$ with
$x_n=\r{{2n}}/\r{2}^n$.
%and $x_4=\r{8}/\r{2}^4$, respectively.
These ratios were taken from the results of the global fit to all data.

%fffffffffffffffffffffffffffffffffffffffffffffffffffffffffffffffffffffffffff
\begin{figure}[tbh]
\includegraphics[width=7.5cm]{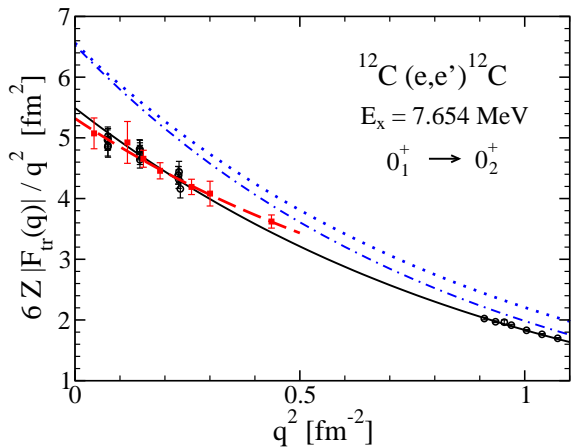}
\caption{ (color online). Experimental form factors at low $q$ corrected
by Eq.~(\ref{eq:FF-exp}) compared to different model approaches.
Extrapolation to zero momentum transfer provides the monopole matrix element.
Square symbols and dashed line: new data and polynomial fit with Eq.~(\ref{eq:fit}).
Circles and full line: old data and global fit with
Eq.~(\ref{eq:FFfit}).
Dashed-dotted line: FMD model, dotted line: $\alpha$-cluster model
\cite{ref:Chernykh2007}.
\label{fig:MatrixElementExtraction}}
\end{figure}
%ffffffffffffffffffffffffffffffffffffffffffffffffffffffffffffffffffffffffffff

The dashed curve  in Fig.~\ref{fig:MatrixElementExtraction} represents
the fit to the data resulting in $\r{2}^\lowq$~=~$(5.29 \pm
0.14)~$fm$^2$ and $\r{4}^\lowq$~=~$(96 \pm 10)~$fm$^4$.
%The result depends very little on the ratios $x_3$ and $x_4$. By changing
%$x_3=14.3$ and $x_4=54.6$ from the global fit to $x_3=11.3$ and
%$x_4=45.7$, which is the FMD result, the deduced transition matrix
%elements stay well within their error bars.
As can be seen from Fig.~\ref{fig:MatrixElementExtraction} an
extraction of the transition radius by extrapolation to $q = 0$
considering only low-$q$ data is very sensitive to the experimental
errors. The global fit to all data (full line) reproduces the new
low-$q$ data except for a single data point at $q^2=0.437$~fm$^{-2}$.
The slope of the polynomial fit (dashed line) is apparently not steep
enough to match the data at higher momentum transfer around
$q^2=1$~fm$^{-2}$. The inherent uncertainty when considering low-$q$
data only has also been pointed out by Sick for an extraction of the
proton charge radius from elastic electron scattering
\cite{ref:Sick2003}. Although the value of $\Gamma_\pi$ obtained with
our new data at low $q$ agrees within error bars with the global fit it
is clear from Fig.~\ref{fig:MatrixElementExtraction} that the fit of
all available data with a suitable ansatz is more reliable.

The deviations of the model transition densities at large radii (cf.\
Fig.~\ref{fig:rhofit}) lead to $\r{2}$ values exceeding the
experimental results. Moreover, Fig.~\ref{fig:MatrixElementExtraction}
demonstrates that also the slopes at $q=0$ (proportional to $\r{4}$)
are too steep indicating the need for an improved description of the
tails of the model wave functions.

% SUMMARY

To summarize, the pair decay width of the Hoyle state has been
extracted from a new $^{12}$C(e,e$^{\prime}$) experiment at low
momentum transfers and independently with a novel model-independent
global fit of the world data. The latter, shown to be less prone to
systematic errors, provides $\Gamma_{\pi} = (62.3\pm 2.0)$~$\mu$eV in
accord with values deduced 40 years ago but with much reduced
uncertainty.
%To summarize, we performed a new $^{12}$C(e,e$^{\prime}$) experiment on
%the Hoyle state at low momentum transfers and extracted with a novel
%model-independent global fit a pair width $\Gamma_{\pi} = (62.3\pm
%2.0)$~$\mu$eV in accord with pair widths deduced 40 years ago but with
%much reduced uncertainty.
%
Combined with an improved determination of the pair width branching
ratio presently underway, the astrophysical reaction rate of the
triple-$\alpha$ process is then known with a precision sufficient to
quantitatively constrain the modeling of a variety of key problems in
nuclear astrophysics.

The result for the pair width deviates from a more recent Fourier
Bessel (FB) analysis. The origin of this discrepancy is explained by
shortcomings of the FB and the lack of data beyond $q=3.1$~fm$^{-1}$.
Furthermore, the study of the monopole transition to the Hoyle state by
electron scattering at low momentum transfer provides important
constraints on models by a unique test of the nuclear wave function at
large distances where one expects $\alpha$-cluster structures.

% ACKNOWLEGEMENTS

H.-D.~Gr\"af, R.~Eichhorn and the S-DALINAC team are thanked for
preparing excellent electron beams. Discussions with S.~M.~Austin are
gratefully acknowledged. We are indebted to H.~P.~Blok for his help
with the DWBA code and to H.~Crannell for detailed information on the
previous FB analysis. This work has been supported by the DFG under
contract SFB 634.


\begin{thebibliography}{abc99x}

\bibitem{ref:Fynbo2005}
H. O. U. Fynbo {\it et al.},
%H. O. U. Fynbo, C. A. Diget, U. C. Bergmann, M. J. G. Borge,
%J. Cederk\"all, P. Dendooven, L. M. Fraile, S. Franchoo,
%V. N. Fedosseev, B. R. Fulton, W. Huang, J. Huikari, H. B. Jeppesen,
%A. S. Jokinen, P. Jones, B. Jonson, U. K\"oster, K. Langanke,
%M. Meister, T. Nilsson, G. Nyman, Y. Prezado, K. Riisager,
%S. Rinta-Antila, O. Tengblad, M. Turrion, Y. Wang, L. Weissman,
%K. Wilhelmsen, J. \"Ayst\"o, and the ISOLDE Collaboration,
%
Nature {\bf 433}, 136 (2005).


\bibitem{ref:Wallerstein1997}
G. Wallerstein {\it et al.},
%
%G. Wallerstein, I. Iben, P. Parker, A. M. Boesgaard, G. M. Hale, A.
%E. Champagne, C. A. Barnes,F. K\"appeler, V. V. Smith, R. D.
%Hoffman, F. X. Timmes, C. Sneden, Chris, R. N. Boyd, B. S. Meyer,
%and D. L. Lambert,
%
Rev. Mod. Phys. {\bf 69}, 995 (1997).

\bibitem{ref:Austin2005} S. M.~Austin,
%
Nucl. Phys. {\bf A758}, 375c (2005).

\bibitem{ref:Tur2007}
C. Tur, A. Heger, and S. M. Austin,
%
Ap. J. {\bf 671}, 821 (2007).

\bibitem{ref:Herwig2006}
F. Herwig, S. M. Austin, and J. C. Lattanzio,
%
Phys. Rev. C {\bf 73}, 025802 (2006).

\bibitem{ref:Kappeler1998}
%
F. K\"appeler, F.-K. Thielemann, and M. Wiescher,
%
Annu. Rev. Nucl. Part. Sci. {\bf 48}, 175 (1998).

\bibitem{ref:Tur2009} C. Tur, A. Heger, and S. M. Austin,
%
Ap. J. {\bf 702}, 1068 (2009).

\bibitem{ref:Hoyle1954}
F. Hoyle,
%
Ap. J. Suppl. {\bf 1}, 121 (1954).

\bibitem{ref:Markham1976}
R. G. Markham, S. M. Austin, and M. A. M. Shahabuddin,
%
Nucl. Phys. {\bf{A270}}, 489 (1976).

\bibitem{ref:Alburger1977}
D. E.~Alburger,
%
Phys. Rev. C {\bf{16}}, 2394 (1977).

\bibitem{ref:Austinprivcomm}
S. M. Austin,
%
private communication.

\bibitem{ref:Wilkinson1969} D. H.~Wilkinson,
%
Nucl.~Phys. {\bf{A133}}, 1 (1969).

\bibitem{ref:kibedi2005}
T. Kib\'{e}di and R. H. Spear,
%
At. Data Nucl. Data Tables {\bf 89}, 77 (2005).

\bibitem{ref:Crannell1967}
H. Crannell {\it et al.},
%H.~Crannell, T.A.~Griffy, L.R.~Suelzle, and M.R.~Yearian,
%
Nucl. Phys. {\bf{A90}}, 152 (1967).

\bibitem{ref:Strehl1970}
P.~Strehl,
%
Z. Phys. {\bf{234}}, 416 (1970).

\bibitem{ref:Ajzenberg-Selove1990}
F.~Ajzenberg-Selove,
%
Nucl. Phys. {\bf{A506}}, 79 (1990).

%\bibitem{ref:Dreher1974}
%B. Dreher {\it et al.},
%B.~Dreher, J.~Friedrich, K.~Merle, H.~Rothhaas, and G.~L\"uhrs,
%
%Nucl. Phys. {\bf{A235}}, 219 (1974).

\bibitem{ref:Heisenberg1981}
J.~Heisenberg,
%
Adv.~Nucl.~Phys. {\bf{12}}, 61 (1981).

\bibitem{ref:Crannell2005}
H. Crannell {\it et al.},
%H.~Crannell, X.~Jiang, J. T.~O'Brien, D. I.~Sober, and E.~Offermann,
%
Nucl. Phys. {\bf{A758}}, 399c (2005).

\bibitem{ref:Chernykh2007}
M. Chernykh {\it et al.},
%M.~Chernykh, H.~Feldmeier, T.~Neff, P.~von~Neumann-Cosel, and A. Richter
%
Phys. Rev. Lett. {\bf{98}}, 032501 (2007).

\bibitem{ref:Tohsaki2001}
A. Tohsaki {\it et al.},
%A.~Tohsaki, H.~Horiuchi, P.~Schuck, and G.~R\"opke,
%
Phys. Rev. Lett. {\bf{87}}, 192501 (2001).

\bibitem{ref:lenhardt2006}
A. W. Lenhardt {\it et al.},
%A. W.~Lenhardt, U.~Bonnes, O.~Burda, P.~von~Neumann-Cosel,
%M.~Platz, A.~Richter, and S.~Watzlawik,
%
Nucl. Instrum. Methods in Phys. Research A {\bf{562}}, 320 (2006).

\bibitem{ref:Heisenberg1983}
J.~Heisenberg and H. P.~Blok,
%
Annu. Rev. Nucl. Part. Sci. {\bf{33}}, 569 (1983).

\bibitem{ref:Roth2004}
R. Roth, T. Neff, H. Hergert, and H. Feldmeier,
%
Nucl. Phys {\bf A745}, 3 (2004).

\bibitem{ref:Sick2003}
I.~Sick,
Phys. Lett. B {\bf 576}, 62 (2003).

\end{thebibliography}
\end{document}